\documentclass[review, authoryear]{elsarticle}

\usepackage{booktabs}
\usepackage{array}
\usepackage{amsmath, amsthm}
\usepackage{amssymb}
\usepackage{url}
\usepackage{natbib}
\usepackage{amsfonts}
\usepackage{microtype}
\usepackage[multiple]{footmisc}

\usepackage{todonotes} 

\usepackage[primary]{xqting}

\newcommand{\citeeg}[1]{\citep[e.g.,][]{#1}}

\DeclareUrlCommand\DOI{}
\newcommand{\doi}[1]{\href{http://dx.doi.org/#1}{\DOI{doi:#1}}}



\renewcommand{\|}{\,|\,}

\DeclareMathOperator*{\argmax}{arg\,max}

\usepackage{dsfont}
\newcommand{\N}{\mathds{N}}

\newcommand{\D}{\mathcal{D}}

\providecommand{\defnqed}{\hfill\ensuremath{\qed}}
\newcommand{\noun}[1]{\textsc{#1}}

\theoremstyle{definition}
\newtheorem{definition}{Definition}

\begin{document}

\title{Predicting Human Behavior in\\Unrepeated, Simultaneous-Move Games\tnoteref{t1}}
\tnotetext[t1]{Preliminary versions of portions of this work appeared in the proceedings of two computer science conferences \citep{wright10, wright12}.}

\author{James R.\ Wright\corref{cor1}}
\ead{jrwright@cs.ubc.ca}

\author{Kevin Leyton-Brown\corref{cor2}}
\ead{kevinlb@cs.ubc.ca}

\address{Department of Computer Science,
201-2366 Main Mall,
University of British Columbia,
Vancouver, BC, Canada,
V6T 1Z4}

\cortext[cor1]{Principal corresponding author, Telephone +1 917 472 8253}
\cortext[cor2]{Corresponding author, Telephone +1 604 822 1453}

\begin{abstract}
  It is common to assume that agents will adopt Nash equilibrium strategies; however, experimental studies have demonstrated that Nash equilibrium is often a poor description of human players' behavior in unrepeated normal-form games.
  %
  In this paper, we analyze five widely studied models (Quantal Response Equilibrium, Level-$k$, Cognitive Hierarchy, QLk, and Noisy Introspection) that aim to describe actual, rather than idealized, human behavior in such games.  We performed what we believe is the most comprehensive meta-analysis of these models, leveraging ten different data sets from the literature recording human play of two-player games.  We began by evaluating the models' \emph{generalization} or \emph{predictive} performance, asking how well a model fits unseen \emph{test data} after having had its parameters calibrated based on separate \emph{training data}.
  Surprisingly, we found that what we dub the QLk model of \cite{stahl94} consistently achieved the best performance. Motivated by this finding, we describe methods for analyzing the posterior distributions over a model's parameters. We found that QLk's parameters were being set to values that were not consistent with their intended economic interpretations. We thus explored variations of QLk, ultimately identifying a new model family that has fewer parameters, gives rise to more parsimonious parameter values, and achieves better predictive performance.
\end{abstract}


\maketitle

\section{Introduction}

In strategic settings, it is common to assume that agents will adopt Nash equilibrium strategies, behaving so that each optimally responds to the others.  This solution concept has many appealing properties; e.g., under any other strategy profile, one or more agents will regret their strategy choices. However, experimental evidence shows that Nash equilibrium often fails to describe human strategic behavior \citep[see, e.g.,][]{goeree01}---even among professional game theorists \citep{becker05}.

%
The relatively new field of \emph{behavioral game theory} extends game-theoretic models to account for human cognitive biases and limitations \citep{camerer03}.  Experimental evidence is the foundation of behavioral game theory, and researchers have developed many models of how humans behave in strategic situations based on such data.
%
This multitude of models presents a practical problem, however: which should we use to predict human behavior?
%
Existing work in behavioral game theory does not directly answer this question, for two reasons.  First, it has tended to focus on explaining (fitting) in-sample behavior rather than predicting out-of-sample behavior.  This means that models are vulnerable to \emph{overfitting} the data: the most flexible model can be chosen instead of the most accurate one.  Second, behavioral game theory has tended not to compare multiple behavioral models, instead either exploring elaborations of a single model or comparing only to one other model (typically Nash equilibrium).
In this work we perform rigorous---albeit computationally intensive---comparisons of many different models and model variations on a wide range of experimental data, leading us to believe that ours is the most comprehensive study of its kind.

Our focus is on the most basic of strategic interactions: unrepeated (``initial'') play in simultaneous move games. In the behavioral game theory literature, five key paradigms have emerged for modeling human decision making in this setting: quantal response equilibrium \citep[QRE;][]{mckelvey95}; the noisy introspection model \citep[NI;][]{goeree04}; the cognitive hierarchy model \citep[CH;][]{camerer04cognitive}; the closely related level-$k$ \citep[Lk;][]{costagomes01,nagel95} models; and what we dub quantal level-$k$ \citep[QLk;][]{stahl94} models.  Although there exist studies exploring different variations of these models \citep[e.g.,][]{stahl95,ho98,weizsacker03,rogers09}, the overwhelming majority of behavioral models of initial play of normal-form games fall broadly into this categorization.  

The first contribution of our work is methodological: we demonstrate broadly applicable techniques for comparing and analyzing behavioral models. (See Section~\ref{sec:recommendations} for our specific methodological recommendations.) We illustrate the use of these techniques via an extensive meta-analysis based on data published in ten different studies, rigorously comparing Lk, QLk, CH, NI, and QRE to each other and to a model based on Nash equilibrium.
The findings that result from this meta-analysis both demonstrate the usefulness of the approach and constitute our second contribution.
Our first main finding is that QLk is the best performing of these predictive models, both on most individual source datasets and also on a dataset pooling all of the ten datasets.
We then analyze and interpret the parameter distributions for several models, including QLk.
Based on this analysis, we construct and evaluate a family of variations on QLk.
Our second main finding is that a simpler (two-parameter) model achieves better out-of-sample predictive performance than any of the models from the literature that we considered.
We recommend the use of this model, dubbed Poisson-QCH, by researchers wanting to predict human play in unrepeated normal-form games.

All of the models we consider depend upon exogenous parameters. 
Most previous work has focused on models' ability to \emph{describe} human behavior, and hence has sought parameter values that best explain  observed experimental data, or more formally that maximize a dataset's probability.\footnote{All of the models that we consider make probabilistic predictions; thus, we must score models according to how much probability mass they assign to observed events, rather than assessing accuracy.} We depart from this descriptive focus, seeking to find models, and hence parameter values, that are effective for \emph{predicting} previously unseen human behavior. Thus, we follow a different approach taken from machine learning and statistics. We begin by randomly dividing the experimental data into a training set and a test set.  We then set each model's parameters to values that maximize the likelihood of the training dataset, and finally score the each model according to the disjoint test dataset's likelihood.  To reduce the variance of this estimate without biasing its expected value, we employ cross-validation \mbox{\citep[see, e.g.,][]{bishop06},} systematically repeating this procedure with different test and training sets.

Our meta-analysis has led us to draw three qualitative conclusions.  First, and least surprisingly, Nash equilibrium is less able to explain human play than are behavioral models.  Second, two high-level themes that underlie the five behavioral models, which we dub ``cost-proportional errors'' and ``limited iterative strategic thinking'', appear to model independent phenomena. Third, and building on the previous conclusion, 
%
the quantal level-$k$ model of \citet{stahl94} (QLk)---which combines both of these themes---made the most accurate predictions. Specifically, QLk substantially outperformed all other models on a new dataset spanning all data in our possession, and also had the best or nearly the best performance on each individual dataset.
%
Our findings were quite robust to variation in the games played by human subjects.  We broke down model performance by game properties such as dominance structure and number/types of equilibria, and obtained essentially the same results as on the combined dataset.
%
We do note that our datasets consisted entirely of two-player games.
Previous work suggests that human subjects reason about $n$-player games as if they were two-player games, failing to fully account for the independence of the other players' actions \citep{ho98, costagomes09}; we might thus expect to observe qualitatively similar results in the $n$-player case.
Nevertheless, empirically confirming this expectation is an important future direction.

The approach we have described so far is designed to compare model performance,  but yields little insight into how or why a model works. For example, maximum likelihood estimates provide no information about the extent to which parameter values can be changed without a large drop in predictive accuracy, or even about the extent to which individual parameters influence a model's performance.  We thus introduce an alternate, Bayesian approach for gaining understanding about a behavioral model's entire parameter space.  We combine experimental data with explicitly quantified prior beliefs to derive a posterior distribution that assigns probability to parameter settings in proportion to their consistency with the data and the prior \citep{gill02}.
%
Applying this approach, we analyze the posterior distributions for three models: a model based on Nash equilibrium, QLk, and Poisson--Cognitive Hierarchy (Poisson-CH). Although Poisson-CH did not demonstrate competitive performance in our initial model comparisons, we analyze it because it is one-dimensional and because of a very concrete and influential recommendation in the literature: \citet{camerer04cognitive} recommended setting the model's single parameter, which represents agents' mean number of steps of strategic reasoning, to $1.5$. Our own analysis sharply contradicts this recommendation, placing the 99\% confidence interval almost a factor of three lower, on the range $[0.51,0.59]$.  We devote most of our attention to QLk, however, due to its extremely strong performance.
%
Our new analysis points out multiple anomalies in QLk's optimal parameter settings, suggesting that a simpler model could be preferable. We thus exhaustively evaluated a family of variations on QLk, thereby identifying a simpler, more predictive family of models based in part on the cognitive hierarchy concept.  In particular, we introduce a new three-parameter model that gives rise to a more plausible posterior distribution over parameter values, while also achieving better predictive performance than five-parameter QLk.

In the next section, we define the models that we study.  Section~\ref{sec:methods1} lays out the formal framework within which we work, and Section~\ref{sec:setup} describes our data, methods, and the Nash-equilibrium-based model to which we compare the behavioral models.  Section~\ref{sec:comparisons} presents the results of our comparisons. Section~\ref{sec:methods2} introduces our methods for Bayesian parameter analysis, and Section~\ref{sec:parameters} describes the anomalies we identified by applying this analysis to our datasets.  Section~\ref{sec:variations} explains the space of QLk variations that we investigated, and introduces our new, high-performing three-parameter model.  In Section~\ref{sec:perf-related} we survey related work from the literature and explain how our own work contributes to it. We conclude in Section~\ref{sec:conclusions}. We defer derivations to appendices. A final appendix investigates the sensitivity of our results to dataset composition, studying how model performance varies with important game properties such as degree of dominance solvability and Nash equilibrium structure. 

\section{Models for Predicting Human Play of Simul\-taneous-Move Games}
\label{sec:existing}
Formally, a behavioral model is a mapping from a game description $G$ and a vector of parameters $\theta$ to a predicted distribution over each action profile $a$ in $G$, which we denote $\Pr(a \| G, \theta)$.  In what follows, we define five prominent behavioral models of human play in unrepeated, simultaneous-move games.\footnote{We focus here on models of behavior in general one-shot, normal-form games.  We omit models of learning in repeated normal-form games such as impulse-balance equilibrium \citep{selten94}, payoff-sampling equilibrium \citep{osborne98}, action-sampling equilibrium \citep{selten08}, and experience-weighted attraction \citep{camerer99}, and models restricted to single game classes, 
such as cooperative equilibrium \citep{capraro13}.  We also omit variants and generalizations of the models we study, such as those introduced by \citet{rogers09}, \citet{weizsacker03}, and \citet{cabrera07}; however, see Section~\ref{sec:variations}, where we systematically explored a particular space of variants.}

\subsection{Quantal Response Equilibrium}
\label{sec:qre}
One important idea from behavioral economics is that people become more likely to make errors as those errors become less costly; we call this making \emph{cost-proportional errors}.  This can be modeled by assuming that agents best respond \emph{quantally}, rather than via strict maximization.
\begin{definition}[Quantal best response]
  Let $u_i(a_i,s_{-i})$ be agent $i$'s expected utility in game $G$ when playing action $a_i$ against strategy profile $s_{-i}$.  Then a \emph{(logit) quantal best response} $QBR^G_i(s_{-i}; \lambda)$ by agent $i$ to $s_{-i}$ is a mixed strategy $s_i$ such that
  \begin{equation}
    s_i(a_i) =  \frac{\exp[\lambda\cdot u_i(a_i, s_{-i})]}{\sum_{a'_i}\exp[\lambda \cdot u_i(a'_i, s_{-i})]},\label{eq:qbr}
  \end{equation}
  where $\lambda$ (the \emph{precision} parameter) indicates how sensitive agents are to utility differences, with $\lambda=0$ corresponding to uniform randomization and $\lambda\rightarrow\infty$ corresponding to best response.
  When its value is clear from context, we will omit the precision parameter.
  Note that unlike best response, which is a set-valued function, quantal best response always returns a unique mixed strategy.\defnqed
\end{definition}
The notion of quantal best response gives rise to a generalization of Nash equilibrium known as the \emph{quantal response equilibrium} (``QRE'') \citep{mckelvey95}.
\begin{definition}[QRE]
  A \emph{quantal response equilibrium} with precision
  $\lambda$ is a mixed strategy profile $s^*$ in which every agent's
  strategy is a quantal best response to the strategies of the other
  agents; i.e., $s^*_i = QBR^G_i(s^*_{-i};\lambda)$ for all agents $i$.\defnqed
\end{definition}

A QRE is guaranteed to exist for any normal-form game and non-negative precision \citep{mckelvey95}.  However, QRE are not guaranteed to be unique.  As is standard in the literature, we select the (unique) QRE that lies on the principal branch of the QRE homotopy at the specified precision.  The principal branch has the attractive feature of approaching the risk-dominant equilibrium as $\lambda\to\infty$ in $2 \times 2$ games with two strict equilibria \citep{turocy05}.

Although Equation \eqref{eq:qbr} is translation invariant, it is not scale invariant.
That is, while adding some constant value to the payoffs of a game will not change its QRE, multiplying payoffs by a positive constant will.
This is problematic because utility functions are only unique up to affine transformations \citep{vonneumann44}; 
hence, equivalent utility functions that have been multiplied by different constants will induce different QREs.
The QRE concept nevertheless makes sense if human players are believed to play games differently depending on the magnitudes of the payoffs involved.

\subsection{Level-k}

Another key idea from behavioral economics is that humans can perform only a limited number of \emph{iterations of strategic reasoning}.
The level-$k$ model \citep{costagomes01} captures this idea by associating each agent $i$ with a level $k_i \in \{0, 1, 2, \ldots\}$, corresponding to the number of iterations of reasoning the agent is able to perform.
A \emph{level-$0$ agent} plays randomly, choosing uniformly at random from his possible actions.  A \emph{level-$k$ agent}, for $k \ge 1$, best responds to the strategy played by level-$(k-1)$ agents.  If a level-$k$ agent has more than one best response, he mixes uniformly over them.

We consider a particular level-$k$ model, dubbed Lk, which assumes that all agents belong to levels 0,1, and 2.\footnote{\label{fn:lk}
We here model only level-$k$ agents, unlike \citet{costagomes01} who also modeled other decision rules.  Like \citet{costagomes01}, we restrict agents' levels to be no greater than 2; however, see Section~\ref{sec:variations}, in which we extend this level-$k$ model to higher levels.} Each agent with level $k>0$ has an associated probability $\epsilon_k$ of making an ``error'', i.e., of playing an action that is not a best response to the level-$(k-1)$ strategy.  Agents are assumed not to account for these errors when forming their beliefs about how lower-level agents will act.
\begin{definition}[Lk model]
  Let $A_i$ denote player $i$'s action set and let $BR^G_i(s_{-i})$ denote the set of $i$'s best responses in game $G$ to the strategy profile $s_{-i}$.  Let $IBR^G_{i,k}$ denote the \emph{iterative best response set} for a level-$k$ agent $i$, with $IBR^G_{i,0}=A_i$ and $IBR^G_{i,k} = BR^G_i(IBR^G_{-i,k-1})$.  Then the distribution $\pi^{Lk}_{i,k} \in \Pi(A_i)$ that the Lk model predicts for a level-$k$ agent $i$ is defined as
  \begin{align*}
    \pi^{Lk}_{i,0}(a_i) &= |A_i|^{-1}, \\
    \pi^{Lk}_{i,k}(a_i) &=
    \begin{cases}
      (1-\epsilon_k)/|IBR^G_{i,k}| &\text{ if } a_i \in IBR^G_{i,k},\\
      \epsilon_k/(|A_i| - |IBR^G_{i,k}|) &\text{otherwise.}
    \end{cases}
  \end{align*}
  The overall predicted distribution of actions is a weighted sum of the distributions for each level:
  \[\Pr(a_i \| G, \alpha_1, \alpha_2, \epsilon_1, \epsilon_2) = \sum_{\ell=0}^2 \alpha_\ell\cdot \pi^{Lk}_{i,\ell}(a_i),\]
  where $\alpha_0 = 1 - \alpha_1 - \alpha_2$. This model thus has 4 parameters: $\{\alpha_1, \alpha_2\}$, the proportions of level-$1$ and level-$2$ agents, and $\{\epsilon_1, \epsilon_2\}$, the error probabilities for level-$1$ and level-$2$ agents.\defnqed
\end{definition}

\subsection{Cognitive Hierarchy}
\label{sec:poisson-ch}
The cognitive hierarchy model \citep{camerer04cognitive}, like level-$k$, models agents with heterogeneous bounds on iterated reasoning.  It differs from the level-$k$ model in two ways.  First, according to this model agents do not make errors; each agent always best responds to its beliefs.  Second, agents of level-$m$ best respond to the full distribution of agents at levels $0$ to $(m-1)$, rather than only to level-$(m-1)$ agents.  More formally, every agent has an associated level $m \in \{0, 1, 2, \ldots\}$.  Let $f$ be a probability mass function describing the distribution of the levels in the population.  Level-$0$ agents play uniformly at random.  Level-$m$ agents ($m \ge 1$) best respond to the strategies that would be played in a population described by the truncated probability mass function $f(j \| j<m)$.

\citet{camerer04cognitive} advocate a single-parameter restriction of the cognitive hierarchy model called \emph{Poisson-CH}, in which $f$ is a Poisson distribution.
\begin{definition}[Poisson-CH model]
  Let $\pi^{PCH}_{i,m} \in \Pi(A_i)$ be the distribution over actions predicted for an agent $i$ with level $m$ by the Poisson-CH model.  Let $f(m) = \text{Poisson}(m;\tau)$.  Let $BR^G_i(s_{-i})$ denote the set of $i$'s best responses in game $G$ to the strategy profile $s_{-i}$.  Let \[\pi^{PCH}_{i,0:m}=\sum_{\ell=0}^mf(\ell)\frac{\pi^{PCH}_{i,\ell}}{\sum_{\ell'=0}^mf(\ell')}\] be the truncated distribution over actions predicted for an agent conditional on that agent's having level $0 \le \ell \le m$.  Then $\pi^{PCH}$ is defined as
  \begin{align*}
    \pi^{PCH}_{i,0}(a_i) &= |A_i|^{-1}, \\
    \pi^{PCH}_{i,m}(a_i) &=
    \begin{cases}
      |BR^G_i(\pi^{PCH}_{i,0:m-1})|^{-1}  &\text{ if } a_i \in BR^G_i(\pi^{PCH}_{i,0:m-1}), \\
      0                              &\text{ otherwise.}
    \end{cases}
  \end{align*}
  The overall predicted distribution of actions is a weighted sum of the distributions for each level,
  \[\Pr(a_i \| G, \tau) = \sum_{\ell=0}^\infty f(\ell)\cdot \pi^{PCH}_{i,\ell}(a_i).\]
 The Poisson distribution's mean, $\tau$, is thus this model's single parameter.\defnqed
\end{definition}

\citet{rogers09} note that cognitive hierarchy and QRE often make similar predictions.  One possible explanation for this is that cost-proportional errors are adequately captured by cognitive hierarchy (and other iterative models), even though they do not explicitly model this effect.  Alternatively, these phenomena could be sufficiently distinct that explicitly modeling both limited iterative strategic thinking and cost-proportional errors yields improved predictions.

\subsection{Quantal Level-k}

\citet{stahl94} propose a rich model of strategic reasoning that combines elements of the QRE and level-$k$ models; we refer to it as the QLk model (for quantal level-$k$).  In QLk, agents have one of three levels, as in Lk.\footnote{\citet{stahl94} also consider an extended version of this model that adds a type that plays the equilibrium strategy.  In order to avoid the complication of having to specify an equilibrium selection rule, we do not consider this extension, as many of the games in our dataset have multiple equilibria.  See Section~\ref{sec:nash-model} for bounds on the performance of Nash equilibrium predictions on our dataset.}
Each agent responds to its beliefs quantally, as in QRE.

A key difference between QLk and Lk is in the error structure.  In Lk, higher-level agents believe that all lower-level agents best respond perfectly, although in fact every agent has some probability of making an error.  In contrast, in QLk, agents are aware of the quantal nature of the lower-level agents' responses, but have (possibly incorrect) beliefs about the lower-level agents' precision.  That is, level-$1$ and level-$2$ agents use potentially different precisions ($\lambda$'s), and furthermore level-$2$ agents' beliefs about level-$1$ agents' precision can be wrong.

\begin{definition}[QLk model]
The probability distribution $\pi^{QLk}_{i,k} \in \Pi(A_i)$ over actions that QLk predicts for a level-$k$ agent $i$ is
\begin{align*}
  \pi^{QLk}_{i,0}(a_i) &= |A_i|^{-1}, \\
  \pi^{QLk}_{i,1} &= QBR^G_i(\pi^{QLk}_{-i,0} ; \lambda_1), \\
  \pi^{QLk}_{i,1(2)} &= QBR^G_i(\pi^{QLk}_{-i,0} ; \lambda_{1(2)}), \\
  \pi^{QLk}_{i,2} &= QBR^G_i(\pi^{QLk}_{i,1(2)} ; \lambda_2),
\end{align*}
where $\pi^{QLk}_{i,1(2)}$ is a mixed-strategy profile representing level-$2$ agents' prediction of how other agents will play. This can be interpreted either as the level-$2$ agents' beliefs about the behavior of level-$1$ agents alone, or it can be understood as modeling level-$2$ agents' beliefs about both level-$1$ and level-$0$ agents, with the presence of additional level-$0$ agents being captured by a lower precision $\lambda_{1(2)}$.  \citet{stahl94} advocate the latter interpretation.
The overall predicted distribution of actions is the weighted sum of the distributions for each level,
\[\Pr(a_i \| G, \alpha_1,\alpha_2, \lambda_1, \lambda_2, \lambda_{1(2)})
  = \sum_{k=0}^2\alpha_k\pi^{QLk}_{i,k}(a_i),\]
where $\alpha_0=1-\alpha_1-\alpha_2$.
QLk has five parameters: $\{\alpha_1,\alpha_2, \lambda_1, \lambda_2, \lambda_{1(2)}\}$.\defnqed
\end{definition}

\subsection{Noisy Introspection}
\citet{goeree04} propose a model called \emph{noisy introspection} that combines cost-proportional errors and an iterative view of strategic cognition in a different way.  Rather than assuming a fixed limit on the number of iterations of strategic thinking, they instead model cognitive bounds by injecting noise into iterated beliefs about others' beliefs and decisions, with the effect that deeper levels of reasoning are assumed to be noisier.  They then show that this process of noise injection converges to a unique prediction after a finite number of iterations, which for most games is relatively small.

\citeauthor{goeree04} also introduce a concrete version of this model (which we dub NI), in which deeper levels of reasoning are exponentially noisier.  
\begin{definition}[NI model]
    Define $\pi^{NI,n}_{i,k}$ as
    \[
        \pi^{NI,n}_{i,k} = \begin{cases}
                               QBR^G_i(\pi^{NI,n}_{-i,k+1}; \lambda_0/t^k)   &\text{if } k<n, \\
                               QBR^G_i(p_0; \lambda_0/t^n)                &\text{otherwise,}
                           \end{cases}
    \]
    where $p_0$ is an arbitrary mixed profile, $\lambda_0 \ge 0$ is a precision, and $t > 1$ is a ``telescoping'' parameter that determines how quickly noise increases with depth of reasoning.  Then the NI model predicts that each agent will play according to
    \[ \pi^{NI}_i = \lim_{n\to\infty} \pi^{NI,n}_{i,0}.\]  For a fixed game $G$, precision $\lambda_0$, and telescoping parameter $t$, this converges to a unique strategy profile regardless of the choice of $p_0$, since in the limit the precision becomes low enough to bring any profile arbitrarily close to the uniform distribution.\defnqed
\end{definition}

\section{Comparing Models}
\label{sec:methods1}

\subsection{Prediction Framework}
\label{sec:prediction-framework}
How do we determine whether a behavioral model is well supported by experimental data? An experimental dataset
$\D=\{(G_i, \{a_{ij} \| j=1,\ldots,J_i\}) \| i=1,\ldots,I\}$ is a set containing $I$ elements.
Each element is a tuple containing a game $G_i$ and a set of $J_i$ pure actions $a_{ij}$, each played by a human subject in $G_i$. There is no reason to maintain the pairing of the play of a human player with that of his opponent, as games are unrepeated. Recall that a behavioral model is a mapping from a game description $G_i$ and a vector of parameters $\theta$ to a predicted distribution over each action $a_i$ in $G_i$, which we denote $\Pr(a_i \| G_i, \theta)$.  

A behavioral model can only be used to make predictions when its parameters are instantiated. How should we set these parameters? Our goal is a model that produces accurate probability distributions over the actions of human agents, rather than simply determining the single action most likely to be played.  This means that we cannot score different models (or, equivalently, different parameter settings for the same model) using a criterion such as a 0--1 loss function (accuracy), which asks how many actions were accurately predicted. For example, the \mbox{0--1} loss function evaluates models based purely upon which action is assigned the highest probability, and does not take account of the probabilities assigned to the other actions.
%
Instead, we evaluate a given model on a given dataset by \emph{likelihood}.  That is, we compute the probability of the observed actions according to the distribution over actions predicted by the model.
The higher the probability of the actual observations according to the prediction output by a model, the better the model predicted the observations.
This takes account of the full predicted distribution; in particular, for any given observed distribution, the prediction that maximizes the likelihood score is the observed distribution itself.\footnote{Although the likelihood is the quantity that interests us, in practice we operate on the log of the likelihood to avoid numerical precision problems that arise in dealing with exceedingly small quantities. Since log likelihood is a monotonic function of likelihood, a model that has higher likelihood than another model will also have higher log likelihood, and vice versa.}

Assume that there is some true set of parameter values, $\theta^*$, under which the model outputs the true distribution $\Pr(a \| G, \theta^*)$ over action profiles, and that $\theta^*$ is independent of $G$. The maximum likelihood estimate of the parameters based on $\D$,
\[ \hat{\theta} = \argmax_{\theta} \Pr(\D \| \theta),\]
is an unbiased point estimate of the true set of parameters $\theta^*$, whose variance decreases as $I$ grows.
We then use $\hat{\theta}$ to evaluate the model:\footnote{We derive  Equation~\eqref{eq:frequentist} in \ref{apx:likelihood}.}
\begin{equation}
  \Pr(a \| G, \D) = \Pr(a \| G, \hat{\theta}) = \prod_{i=1}^I\prod_{j=1}^{J_i} \Pr(a_{ij} \| G_i, \theta). \label{eq:frequentist}
\end{equation}

\subsection{Assessing Generalization Performance}
\label{sec:methods}

Each of the models that we consider depends on parameters that are estimated from the data.
This presents a problem for evaluating models' performance, since a more flexible model might fit a given dataset better without necessarily predicting unseen data better.  Models that perform well by fitting a specific dataset well, but perform poorly at predicting out-of-sample data (i.e., data that was not used for fitting the model's parameters), are said to \emph{overfit} the data.

There are several approaches to avoiding the overfitting problem.
One is to compare models' fits to the experimental data, but to apply a penalty to models with larger numbers of parameters.  The widely used Bayesian Information Criterion (BIC) and Akaike Information Criterion (AIC) \citeeg{murphy12} take this approach.  However, both criteria are only guaranteed to apply asymptotically in the limit of infinite quantities of data; furthermore, the BIC is only applicable to \emph{nested} models, where one model is a strict generalization of the other.
%
A similar approach is taken by the $\chi$-squared test, which tests the hypothesis that a more-general model's fit is significantly better than that of a restricted model.  However, this is difficult to apply to testing multiple models, in addition to again requiring the models to be nested.
%
A third approach to evaluating predictive performance is to formulate hypotheses based on implications derived directly from a model's definition \citep[see][for examples of such an approach]{haile08,hargreaves14}.  This can be a very effective way of evaluating the predictive performance of a single model; however, due to the binary nature of hypothesis testing, it is less appropriate for comparing multiple models.

In this work, we take a fourth approach, which is widespread in machine learning.  We estimate parameters on a dataset containing a subset of the data (the \emph{training data}), and then evaluate the resulting model by computing likelihood scores on the observations associated with the remaining, disjoint \emph{test data}.  That is, every model's performance is evaluated entirely based on data that were not used for estimating parameters.
We partition data at the level of games: data from a given game appears either in the training set or the test set, but not both.\footnote{
\label{fn:subjects}
\Store{q:subjects}{This means that observations for a given game will appear in exactly one part of the partition.  However, observations from the same subject may appear in multiple parts, when subjects play more than one game.
}}\footnote{
\label{fn:gamewise}In an earlier version of this work, we partitioned our dataset at the level of observations.  Partitioning at the level of games  provides stronger protection against overfitting.
}

Randomly dividing our experimental data into training and test sets introduces variance into the prediction score, since the exact value of the score depends partly upon the random division.
To reduce this variance, we perform 10 rounds of 10-fold \emph{cross-validation}.\footnote{\label{fn:bootstrap}
Repeatedly fitting parameters on a bootstrapped subsample and then evaluating performance on the remaining data is another approach to reducing the variance associated with the division into test and training sets.  This is a more effective approach for reducing the variance of parameter estimates; however, it introduces bias into performance estimates \citep{efron97}, which are our primary focus in this work.}
Specifically, for each round, we randomly partition the games into 10 parts of approximately equal size.  For each of the 10 ways of selecting 9 parts from the 10, we compute the maximum likelihood estimate of the model's parameters based on the observations associated with the games belonging to those 9 parts.  We then determine the likelihood of the observations in the remaining part given the prediction. We call the average of this quantity across all 10 parts the \emph{cross-validated likelihood}.  The average across rounds of the cross-validated likelihoods is distributed according to a Student's-$t$ distribution \citep[see,~e.g.,][]{witten00}.  We compare the predictive power of different behavioral models on a given dataset by comparing the average cross-validated likelihood of the dataset under each model.  We say that one model predicts significantly better than another when the $95\%$ confidence intervals for the average cross-validated likelihoods do not overlap.

\section{Experimental Setup}
\label{sec:setup}
In this section we describe the data and methods that we used in our model evaluations.  We also describe a baseline model based on Nash equilibrium.

\subsection{Data}
\label{sec:data}
As described in detail in Section~\ref{sec:perf-related}, we conducted an extensive survey of papers that make use of the five behavioral models we consider.\footnote{One might wonder whether models tended to do better in datasets from studies that explicitly considered them.  This turned out not to be the case; a given model's performance in a given individual source dataset had essentially no relationship to whether the source dataset had explicitly studied the model.}
We thereby identified ten large-scale, publicly available sets of human-subject experimental data \citep{stahl94,stahl95,costagomes98,goeree01,haruvy01, cooper03, haruvy07, costagomes08, stahl08, rogers09}.  We study all ten\footnote{\label{fn:costagomes06}
We identified an additional dataset \citep{costagomes06} which we do not include due to a computational issue.  The games in this dataset had between $200$ and $800$ actions per player, which made it intractable to compute many solution concepts. As with Nash equilibrium, the main bottleneck in computing behavioral solution concepts is computing expected utilities.  Each epoch of training for this dataset requires calculating expected values over up to $640,000$ outcomes per game, in contrast to between $9$ and approximately $14,000$ outcomes per game in the \noun{All10} dataset.
We attempted to overcome this problem by deriving a coarse version of this data by binning similar actions; however, binning in this way resulted in games that were not strategically equivalent to the originals (e.g., when multiple iterations of best response would result in the same binned action in the coarsened games but different unbinned actions in the original games).
An open problem for future work is finding a way to address this computational problem by representing the games \emph{compactly} \citeeg{kearns01, koller01, jiang11}, such that expected utility can be computed efficiently over even a very large action space.}
of these datasets in this paper.  See Table~\ref{tbl:datasets} for a summary.

\citet{goeree01} presented 10 games in which subjects' behavior was close to that predicted by Nash equilibrium, and 10 other small variations on the same games in which subjects' behavior was \emph{not} well-predicted by Nash equilibrium.  We included the 10 games that were in normal form.
In \citet{cooper03}, agents played the normal forms of 8 games, followed by extensive form games with the same induced normal forms; we include only the data from the normal-form games.
The remaining studies consisted exclusively of normal-form games.

All games had two players, so each single play of a game generated two observations.  We built one dataset for each study.
%
We also constructed a combined dataset, dubbed \noun{All10}, containing data from all the datasets.
The datasets contained very different numbers of observations, ranging from 400 \citep{stahl94} to 2992 \citep{cooper03}.
To ensure that each fold had approximately the same population of subjects, we evaluated \noun{All10} using \emph{stratified} cross-validation: we performed the game partitioning and selection process separately for each of the contained source datasets, thereby ensuring that the number of games from each source dataset was approximately equal in each partition element.

Several studies \citep{stahl94,stahl95,haruvy01,haruvy07,stahl08} paid participants according to a randomized procedure in which experimental subjects played normal-form games for points representing a 1\% chance (per game) of winning a cash prize.
In \citet{costagomes98}, each payoff unit was worth 40~cents, but participants were paid based on the outcome of only one randomly-selected game.
In the remaining studies \citep{goeree01,cooper03,costagomes08,rogers09}, game payoffs were worth a deterministic number of cents.  We summarize the expected value of payoff points in the ``Units'' column of Table~\ref{tbl:datasets}.
The QRE and QLk models depend on a precision parameter that is not scale invariant.  E.g., if $\lambda$ is the correct precision for a game whose payoffs are denominated in cents, then $\lambda/100$ would be the correct precision for a game whose payoffs are denominated in dollars.  To ensure consistent estimation of precision parameters, 
especially in the \noun{All10} dataset where observations from multiple studies were combined, we normalized the payoff values for each game to be in expected cents.

\begin{table}[tb]\centering
  \caption[Names and contents of each dataset.]{
  Names and contents of each dataset.  Units are in expected value, in US dollars.}
  \label{tbl:datasets}
  {
  \begin{tabular}{lm{2in}lll}
      \toprule
        Name         & Source                & Games & $n$ & Units\\
      \midrule       
        \noun{SW94}  & \citet{stahl94}       &  10   &  400  &  \$0.025 \\
        \noun{SW95}  & \citet{stahl95}       &  12   &  576  &  \$0.02  \\
        \noun{CGCB98}& \citet{costagomes98}  &  18   & 1566  &  \$0.022 \\
        \noun{GH01}  & \citet{goeree01}      &  10   &  500  &  \$0.01  \\
        \noun{CVH03} & \citet{cooper03}      &   8   & 2992  &  \$0.10  \\
        \noun{HSW01} & \citet{haruvy01}      &  15   &  869  &  \$0.02  \\
        \noun{HS07}  & \citet{haruvy07}      &  20   & 2940  &  \$0.02  \\
        \noun{CGW08} & \citet{costagomes08}  &  14   & 1792  &  \$0.0107 \\
        \noun{SH08}  & \citet{stahl08}       &  18   & 1288  &  \$0.02  \\
        \noun{RPC08} & \citet{rogers09}      &  17   & 1210  &  \$0.01  \\
      \midrule
        \noun{All10} &  Union of above       & 142   & 13863 &  per source \\
      \bottomrule\\
    \end{tabular}
  }
\end{table}

\subsection{Comparing to Nash Equilibrium}
\label{sec:nash-model}
It is desirable to compare the predictive performance of our behavioral models to that of Nash equilibrium. However, such a comparison is not as simple as one might hope, because any attempt to use Nash equilibrium for prediction must extend the solution concept to address two problems. The first problem is that many games have multiple Nash equilibria; in these cases, the Nash prediction is not well defined. The second problem is that Nash equilibrium frequently assigns probability zero to some actions.
Indeed, in 82\% of the games in our \noun{All10} dataset \emph{every} Nash equilibrium assigned probability 0 to actions that were actually taken by one or more experimental subjects.
This is a problem because we assess the quality of a model by how well it explains the data; unmodified, the Nash equilibrium model considers our experimental data to be \emph{impossible}, and hence receives a likelihood of zero.

We addressed the second problem by augmenting the Nash equilibrium solution concept to say that with some probability, each player chooses an action uniformly at random; this prevents the solution concept from assessing any experimental data as impossible. This probability is a free parameter of the model; as we did with behavioral models, we fit this parameter using maximum likelihood estimation on a training set.
We thus call the model Nash Equilibrium with Error, or NEE.
We sidestepped the first problem by assuming that agents always coordinate to play an equilibrium and by reporting statistics across different equilibria.
Specifically, we report the performance achieved by choosing the equilibrium that respectively best and worst fit the \emph{test} data, thereby giving upper and lower bounds on the test-set performance achievable by any Nash-based prediction. (Note that because we ``cheat'' by choosing equilibria based on test-set performance, these fits are not able to generalize to new data, and hence cannot be used in practice.) Finally, we also reported the prediction performance on the test data, averaged over all of the Nash equilibria of the game.\footnote{One might wonder whether the $\epsilon$-equilibrium solution concept \citep[see e.g.][Section~3.4.7]{shoham08} solves either of these problems.  It does not.
First, $\epsilon$-equilibrium can still assign probability 0 to some actions.
Second, relaxing the equilibrium concept only increases the number of equilibria; indeed, every game has infinitely many $\epsilon$-equilibria for any $\epsilon>0$.  Furthermore, to our knowledge, no algorithm for characterizing this set exists, making equilibrium selection impractical.
}

\subsection{Computational Environment}

We performed computation using WestGrid ({\small\url{www.westgrid.ca}}), primarily on the {\small\texttt{orcinus}} cluster, which has $9600$ 64-bit Intel Xeon CPU cores.
We used {\small\sc{}Gambit} \citep{mckelvey07} to compute QRE and to enumerate the Nash equilibria of games, and computed maximum likelihood estimates using the Nelder--Mead simplex algorithm \citep{nelder65}.

\section{Model Comparisons}
\label{sec:comparisons}
\begin{figure}[t]\centering
  \includegraphics[width=.9\textwidth]{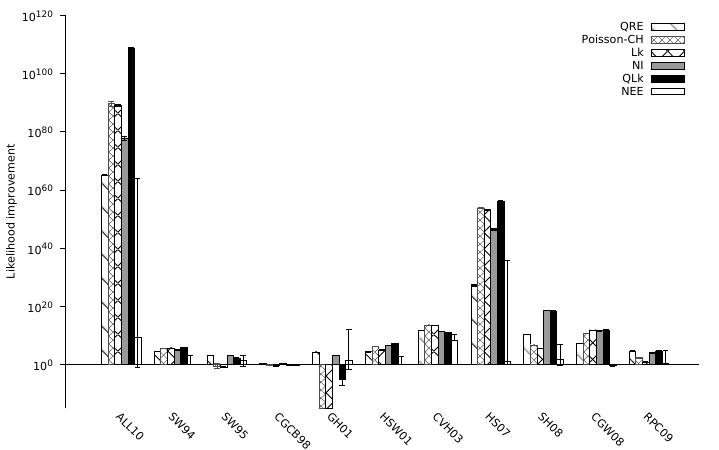}
  \caption[Likelihoods of model predictions]{
  Average likelihood ratios of model predictions to random predictions, with $95\%$ confidence intervals.
  Error bars for NEE show upper and lower bounds on performance depending upon equilibrium selection; the main bar for NEE shows the average performance over all equilibria.  Note that conclusions should not be drawn about relative differences in likelihood across datasets, as likelihood depends on the dataset's number of samples and the underlying games' numbers of actions. Relative differences in likelihood \emph{are} meaningful within datasets.}
  \label{fig:initial}
\end{figure}

In this section we describe the results of our experiments comparing the predictive performance of the five behavioral models from Section~\ref{sec:existing} and of the Nash-based models of Section~\ref{sec:nash-model}.  Figure~\ref{fig:initial} compares our behavioral and Nash-based models. For each model and each dataset, we give the factor by which the dataset was judged more likely according to the model's prediction than it was according to a uniform random prediction.
Thus, for example, the \noun{All10} dataset was approximately $10^{90}$ times more likely to have been generated by an agent acting according to our Poisson-CH model than choosing actions uniformly at random.
For the Nash Equilibrium with Error model, the error bars show the upper and lower bounds on predictive performance obtained by selecting an equilibrium to maximize or minimize test-set performance, and the main bar shows the expected predictive performance of selecting an equilibrium uniformly at random.  For other models, the error bars indicate 95\% confidence intervals across cross-validation partitions; in most cases, these intervals are imperceptibly narrow.

\subsection{Comparing Behavioral Models}
\label{sec:bgt-comparison}

Poisson-CH and Lk achieved very similar performance in most datasets.
In one way this is an intuitive result, since the models are very similar to each other.
On the other hand, it suggests something less obvious, that two differences between the models are not very important in practice: (1) reasoning about just one lower level versus reasoning about the distribution of all lower levels; (2) the distinct error models.

QRE and NI tended to perform well on the same datasets.
On all but two datasets (\noun{HSW01} and \noun{CGW08}), the ordering between QRE and the iterative models was the same as between NI and the iterative models.  We found this result  surprising, since the two models appear quite different. However, the two models do share several key elements in common. First, both models are based around cost-proportional errors, and they both assume that all agents play from the same distribution, unlike the iterative models, which assume that different agents reason to different depths.  Further, although NI is not explicitly a fixed-point model, it does assume an unlimited depth of reasoning, like QRE, although it does typically converge after a relatively small number of iterations.

In five datasets, the models based on cost-proportional errors (QRE and NI) predicted human play significantly better than the two models based on bounded iterated reasoning (Lk and Poisson-CH).
However, in five other datasets, including \noun{All10}, the situation was reversed, with Lk and Poisson-CH outperforming QRE and NI.
In the remaining two datasets, NI outperformed the iterative models, which outperformed QRE.  This mixed result is consistent with earlier, less extensive comparisons of QRE with these two models \citep[see also Section~\ref{sec:perf-related}]{chong05,crawford07fatal,rogers09}, and suggests to us that, in answer to the question posed in Section~\ref{sec:poisson-ch}, there may be value to modeling both bounded iterated reasoning and cost-proportional errors explicitly.
If we were right about this hypothesis, we might expect that our remaining model, which incorporates both components, would predict better than models that are based on only one component.  This was indeed the case: QLk generally outperformed the single-component models.  Overall, QLk was the strongest behavioral model; in a majority of datasets, no model made significantly better predictions.  The datasets in which some model other than QLk did make significantly better predictions were \noun{CVH03}, \noun{SW95}, \noun{CGCB98}, and \noun{GH01}; we discuss the latter in detail below, in Section~\ref{sec:nash-comparison}.

We typically estimated different parameter values than the papers that introduced the models we studied. One reason\footnote{In at least one case, our values are also different due to errors in an original paper's estimation: \citet{stahl94} estimated level proportions that sum to more than 1.} this occurred is that our training set contains a only subset of these games.  This sensitivity to taking subsets of games indicates that overfitting is indeed a realistic concern.

\subsection{Comparing to Nash Equilibrium}
\label{sec:nash-comparison}

It is already widely believed that Nash equilibrium is a poor description of humans' initial play in normal-form games \citeeg{goeree01}.  Nevertheless, for the sake of completeness, we also evaluated the predictive power of Nash equilibrium with error (NEE) on our datasets.  Referring again to Figure~\ref{fig:initial}, we see that NEE's predictions were worse than those of every behavioral model on every dataset except \noun{SW95} and \noun{CGCB98}.  NEE's upper bound---using the post-hoc best equilibrium---was significantly worse than QLk's performance on every dataset except \noun{SW95}, \noun{CGCB98}, \noun{RPC09}, and \noun{GH01}.

\begin{figure}[t]\centering
  \includegraphics{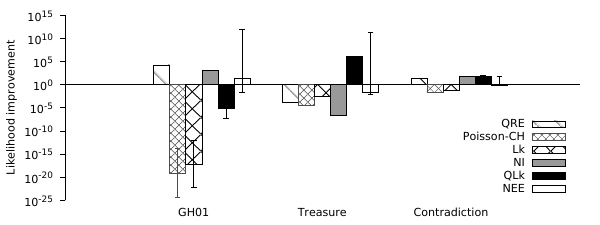}
  \caption[Likelihoods on ``treasure'' and ``contradiction'' treatments]{
    Average likelihood ratios of model predictions to random predictions, with $95\%$ confidence intervals, on \noun{GH01} data separated into ``treasure'' and ``contradiction'' treatments.
    Error bars for NEE show upper and lower bounds on performance depending upon equilibrium selection; the main bar for NEE shows the average performance over all equilibria. Note that relative differences in likelihood are not meaningful across datasets, as likelihood drops with growth in the dataset's number of samples and underlying games' numbers of actions. Relative differences in likelihood \emph{are} meaningful within datasets.}
  \label{fig:gh01-mle}
\end{figure}

NEE's strong performance on \noun{SW95} was surprising; it may have been a result of the unusual subject pool, which consisted of fourth- and fifth-year undergraduate finance and accounting majors.  In contrast, it is unsurprising that NEE performed well on \noun{GH01}, since this distribution was deliberately constructed so that human play on half of its games (the ``treasure'' conditions) would be relatively well described by Nash equilibrium.\footnote{Of course, \noun{GH01} was also constructed so that human play on the other half of its games would be poorly described by Nash equilibrium.
However, this is still a difference from the other datasets, in which Nash equilibrium appears to have poorly described an even larger fraction of games.}
Figure~\ref{fig:gh01-mle} separates \noun{GH01} into its ``treasure'' and ``contradiction'' treatments and compares the performance of the behavioral and Nash-based models on these separated datasets. 
In addition to the fact that the ``treasure'' games were deliberately selected to favor Nash predictions, many of \noun{GH01}'s games have multiple equilibria. This conferred an advantage to our NEE model's upper bound, because it was allowed to pick the equilibrium with best test-set performance on a per-instance basis.
Note that although NEE thus had a higher upper bound than QLk on the  ``treasure'' treatment, its average performance was still quite poor.

\section{Analyzing Model Parameters}
\label{sec:methods2}

Making good predictions from behavioral models depends upon obtaining good estimates of model parameters.  These estimates can also be useful in themselves, helping researchers to understand both how people behave in strategic situations and whether a model's behavior aligns or clashes with its intended economic interpretation. Unfortunately, the method we have used so far---maximum likelihood estimation, i.e., finding a single set of parameters that best explains the training set---is not a good way of gaining this kind of understanding. The problem is that we have no way of knowing how much of a difference it would have made to have set the parameters differently, and hence how important each parameter setting is to the model's performance. If some parameter is completely uncorrelated with predictive accuracy, the maximum likelihood estimate will set it to an arbitrary value, from which we would be wrong to draw economic conclusions.\footnote{We can gain local information about a parameter's importance from the confidence interval around its maximum likelihood estimate: locally important parameters will have narrow confidence intervals, and locally irrelevant parameters will have wide confidence intervals.  However, this does not tell us anything outside the neighborhood of the estimate.}

For example, in the previous chapter we noted that our parameter estimates for QLk implied a much larger proportion of level-$0$ agents than is conventionally expected.  We also interpreted the large estimated value of the noise parameter $\epsilon$ as indicating that Nash equilibrium fits the data poorly.  However, much less can be concluded from such facts if there turn out to be multiple, very different ways of configuring these models to make good predictions.

An alternative is to use Bayesian analysis to estimate the entire posterior distribution over parameter values rather than estimating only a single point.  This allows us to identify the most likely parameter values; how wide a range of values are argued for by the data (equivalently, how strongly the data argues for the most likely values); and whether the values that the data argues for are plausible in terms of our intuitions about parameters' meanings.
We derive an expression for the posterior distribution in \ref{apx:posterior}.
In Section~\ref{sec:parameters} we will apply these methods to study QLk, NEE, and Poisson-CH: the first because it achieved such reliably strong performance;
the second because it has an error term with an especially interpretable posterior distribution;
and the last because it is the model about which the most explicit parameter recommendation was made in the literature.
\citet{camerer04cognitive} recommended setting Poisson-CH's single parameter, which represents agents' mean number of steps of strategic reasoning, to $1.5$.
Our own analysis sharply contradicts this recommendation, placing the 99\% confidence interval roughly a factor of two lower, on the range $[0.70, 0.76]$.
We devote most of our attention to QLk, however, due to its extremely strong performance.

\subsection{Posterior Distribution Estimation}
\label{sec:posterior-estimation}

We estimate the posterior distribution as a set of samples. When a model has a low-dimensional parameter space, like Poisson-CH, we generate a large number of evenly-spaced, discrete points (so-called \emph{grid sampling}). This has the advantage that we are guaranteed to cover the whole space, and hence will not miss large, important regions.
%
However, this approach does not work when a model's parameter space is large, because evenly-spaced grids require a number of samples exponential in the number of parameters. Luckily, we do not care about having good estimates of the whole posterior distribution---what matters is getting good estimates of regions of high probability mass. This can be achieved by sampling parameter settings in proportion to their likelihood, rather than uniformly. A wide variety of techniques exist for performing this sort of sampling.  For models such as QLk with a multidimensional parameter space, we used \emph{Metropolis-Hastings sampling} to estimate the posterior distribution.  The Metropolis-Hastings algorithm is a Markov Chain Monte Carlo (MCMC) algorithm \citeeg{robert04} that computes a series of values from the support of a distribution.
Although each value depends upon the previous value, the values are distributed as if from an independent sample of the distribution after a sufficiently large number of iterations.
MCMC algorithms (and related techniques, e.g., annealed importance sampling \citep{neal01}) are useful for estimating multidimensional distributions for which a closed form of the density is unknown.  They require only that a value \emph{proportional} to the true density be computable (i.e., an unnormalized density).  This is precisely the case with the models that we seek to estimate.

We used a flat prior for all parameters.\footnote{For precision parameters, another natural choice might have been to use a flat prior on the log of precision.
We chose as we did to avoid artificially preferring precision estimates closer to zero, since it is common for iterative models to assume agents best respond nearly perfectly to lower levels.}
Although this prior is improper on unbounded parameters such as precision, it results in a correctly normalized posterior distribution;\footnote{That is, for the posterior, $\idotsint_{-\infty}^\infty\Pr(\theta\|\D)\,d\theta = 1$, even though for the prior $\idotsint_{-\infty}^\infty p_0(\theta)\,d\theta$ diverges.} the posterior distribution in this case reduces to the likelihood \citeeg{gill02}.  For Poisson-CH, where we grid sample an unbounded parameter, we grid sampled within a bounded range ($[0,10]$), which is equivalent to assigning probability $0$ to points outside the bounds.  In practice, this turned out not to matter, as the vast majority of probability mass was concentrated near~$0$.

\subsection{Visualizing Multi-Dimensional Distributions}
In the sections that follow, we present posterior distributions as cumulative marginal distributions.  That is, for every parameter, we plot the cumulative density function (CDF)---the probability that the parameter should be set less than or equal to a given value---averaging over values of all other parameters.  Plotting cumulative density functions allows us to visualize an entire continuous distribution without having to estimate density from discrete samples, thus sparing us manual decisions such as the width of bins for a histogram. Plotting marginal distributions allows us to examine intuitive two-dimensional plots about multi-dimensional distributions.  Interaction effects between parameters are thus obscured; luckily, 
in further, unpublished experiments we found little in the way of interaction effects between parameters.

\section{Parameter Importance Analysis}
\label{sec:parameters}

In this section we analyze the posterior distributions of the parameters for three of the models compared in Section~\ref{sec:comparisons}: Poisson-CH, NEE, and QLk.  We then compare our estimates of the relative proportions of level-$0$ agents to previous work.

For Poisson-CH, we computed the likelihood for each value of $\tau \in \{0.01k \| k\in\N, 0 \le 0.01k \le 10\}$, and then normalized by the sum of the likelihoods.
For NEE, we computed the likelihood for each value of $\epsilon \in \{0.01k \| k\in\N, 0 \le 0.01k \le 1\}$.
%
For Lk and QLk, we combined the samples from $4$ independent Metropolis-Hasting chains, each of which computed $220,000$ samples, discarding the first $20,000$ samples as a ``burn-in'' period to allow the Markov chain to converge.  
We used the PyMC software package to generate the samples \citep{patil10}.  Computing the posterior distribution for a single model in this way typically required approximately 200 CPU hours.

\subsection{Poisson-CH}
\label{sec:pch-results}
\begin{figure}[t]\centering
  \includegraphics{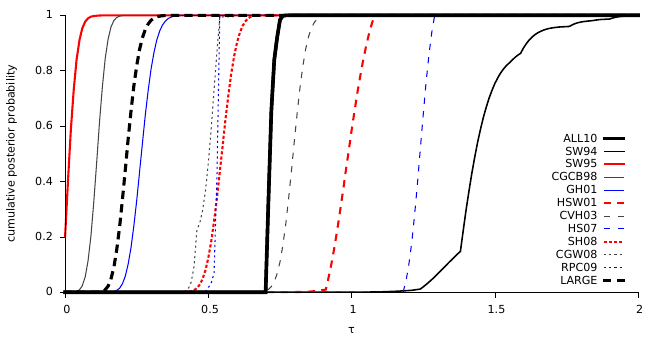}
  \caption[Cumulative posterior distributions for Poisson-CH]{
    Cumulative posterior distributions for Poisson-CH's $\tau$ parameter.
    Bold solid trace is the combined dataset; solid black trace is the outlier \citet{stahl94} source dataset; bold dashed trace is a subset containing all large games (those with more than 5 actions per player).}
  \label{fig:cdf-PCH}
\end{figure}

In an influential recommendation from the literature, \citet{camerer04cognitive} suggest\footnote{
Although \citeauthor{camerer04cognitive} phrase their recommendation as a reasonable ``omnibus guess,'' it is often cited as an authoritative finding \citeeg{carvalho10,frey11,choi12,goodie12}.}
setting the $\tau$ parameter of the Poisson-CH model to $1.5$. Our Bayesian analysis techniques allow us to estimate CDFs for this parameter on each of our datasets (see Figure~\ref{fig:cdf-PCH}). Overall, our analysis strongly contradicts \citeauthor{camerer04cognitive}'s recommendation.
On \noun{All10}, the posterior probability of $0.70 \le \tau \le 0.76$ is more than $99\%$.
%
Every other source dataset had a wider $99\%$ \emph{credible interval} (the Bayesian counterpart to confidence intervals) for $\tau$ than \noun{All10}, as indicated by the higher slope of \noun{All10}'s cumulative density function, since smaller datasets lead to less confident predictions.  Nevertheless, all but two of the source datasets had median values less than $1.0$.  Only the \citet{stahl94} dataset (\noun{SW94}) supports \citeauthor{camerer04cognitive}'s recommendation (median $1.43$).  However, as we have observed before, \noun{SW94} appears to be an outlier; its credible interval is wider than that of the other distributions, and the distribution is very multimodal, possibly due to the dataset's small size.

Many of the games in our dataset have small action spaces.
For example, 108 out of the 142 games in \noun{All10} have exactly 3 actions per player.
One might worry that the estimated average cognitive level in Figure~\ref{fig:cdf-PCH} is artificially low, since it is impossible to distinguish higher numbers of levels than the number of actions available to each player.
We check this by performing the same posterior estimation on a subset of the data consisting only of the 4 large games (i.e., those with more than 5 actions available to each player).
As Figure~\ref{fig:cdf-PCH} shows, the estimated average cognitive level in these large games was even lower than the overall estimate, with a median of $0.22$.

\subsection{Nash Equilibrium}
\label{sec:nee-results}

\begin{figure}[t]\centering
  \includegraphics{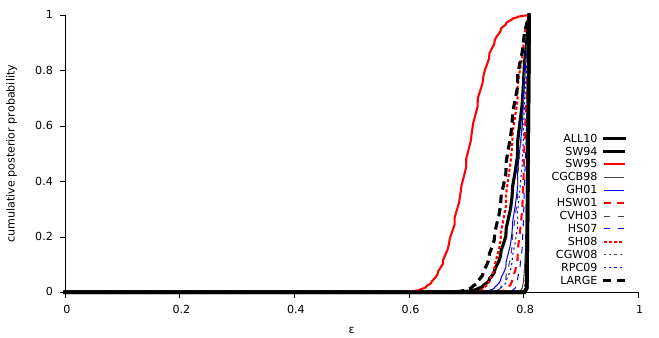}
  \caption[Posterior distributions for NEE]{
    Cumulative posterior distributions for NEE's $\epsilon$ parameter.
    Bold solid trace is the combined dataset; bold dashed trace is a subset containing all large games (those with more than 5 actions per player).}
  \label{fig:cdf-NEE}
\end{figure}

NEE has a free parameter, $\epsilon$, that describes the probability of an agent choosing an action uniformly at random.
If Nash equilibrium were a good tool for predicting human behavior, we would expect this parameter to have a relatively low value; in contrast, the values of $\epsilon$ that maximize NEE's performance were extremely high.
In this section we estimate the full posterior distribution for $\epsilon$; see Figure~\ref{fig:cdf-NEE}.  By doing so we are able to confirm that in both \noun{All10} and its component source datasets, the posterior distribution for $\epsilon$ is very concentrated around very large values of $\epsilon$.
The fact that well over half of NEE's prediction consists of the uniform noise term provides a strong argument against using Nash equilibrium to predict initial play.
This is especially true as the agents within a Nash equilibrium do not take others' noisiness into account, which makes it difficult to interpret $\epsilon$ as a measure of level-$0$ play rather than of model misspecification.

\subsection{QLk}
\label{sec:qlk-results}

\begin{figure}[t]\centering
  \includegraphics{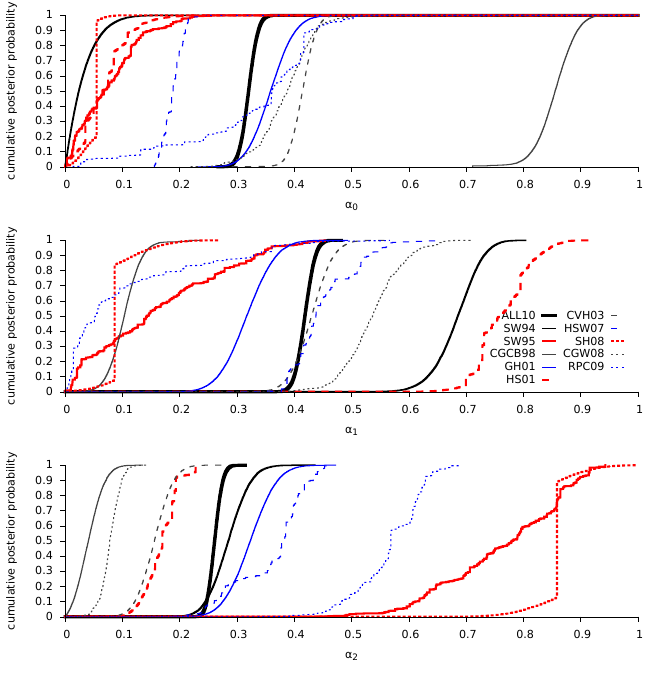}
  \caption{
      Marginal cumulative posterior distribution functions for the level proportion parameters ($\alpha_0, \alpha_1,\alpha_2$) of the QLk model.}
  \label{fig:cdf-qlk-many}
\end{figure}

Figure~\ref{fig:cdf-qlk-many} gives the marginal cumulative posterior distributions for QLk's level proportion distributions broken down by source dataset.  That is, we computed the five-dimensional posterior distribution, and then extracted from it the three marginal distributions shown here.\footnote{\label{fn:lambdas}
We omit marginal distributions for the precision parameters $\lambda_1$, $\lambda_2$, and $\lambda_{1(2)}$ for space reasons.  They follow the same broad pattern as the level proportion distributions: the parameters have relatively diverse posterior distributions and degrees of identification in the individual datasets, but are very sharply identified in the combined \noun{All10} dataset.
}
As with Poisson-CH, posterior level distributions varied across datasets.\footnote{\label{fn:simulation}
To confirm that these results were not simply an artifact of a difficult-to-sample posterior distribution, we simulated data from \noun{All10} from a QLk model with known parameters, and then sampled from the posterior distribution of this synthesized dataset.
For all 5 parameters, the true parameter value was contained within the 95\% central credible interval a minimum of 93 times out of 100 repetitions, indicating that the sampler was well calibrated.
}

We observe a surprisingly high posterior frequency of level-$0$ agents.
The posterior medians for the proportion of level-$0$, level-$1$, and level-$2$ agents in the \noun{All10} dataset are $0.32$, $0.42$, and $0.26$, respectively.
See Section~\ref{sec:level0} for a further discussion of our level-$0$ estimates.

Overall, we observed rather small quantal response precisions.  In the \noun{All10} dataset, the posterior median precisions for level-$1$ agents, level-$2$ agents, and the belief of level-$2$ agents about level-$1$ agents were $0.16$, $0.56$, and $0.05$ respectively.  The belief of the level-$2$ agents that the level-$1$ agents have a much smaller precision than their actual precision was particularly strongly identified.
%
That is, the \noun{All10} dataset assigned the highest posterior probability to parameter settings in which the level-$2$ agents ascribe a smaller than accurate quantal response precision to the level-$1$ agents.
QLk may get this right: e.g., two-level strategic reasoning might cause a high cognitive load, making agents more likely to make mistakes in their predictions of others' behavior. Alternately, we might worry that QLk fails to capture some crucial aspect of experimental subjects' strategic reasoning. For example, the low value of $\lambda_{1(2)}$ might reflect level-$2$ agents' reasoning about all lower levels rather than just one level below themselves: ascribing a low precision to level-$1$ agents approximates a mixture of level-$1$ agents and uniformly randomizing level-$0$ agents.  That is, the low value of $\lambda_{1(2)}$ may be a way of simulating a cognitive hierarchy style of reasoning within a level-$k$ framework.  In the next section, we will explore this possibility as part of an evaluation of systematic variations of QLk's modeling assumptions.

\subsection{Level-0}
\label{sec:level0}

Earlier studies found support for widely varying proportions of level-$0$ agents.
\citet{stahl94} estimated that 0\% of the population was level-$0$;\footnote{Their dataset is an outlier in our own per-dataset parameter fits; see Section~\ref{sec:pch-results}.}
\citet{stahl95} estimated 17\%, with a confidence interval of [6\%, 30\%];
\citet{haruvy01} estimated rates between 6--16\% for various model specifications; and \citet{burchardi14} estimated $37\%$ by fitting a level-$k$ model, and between 20--42\% by eliciting subject strategies.

The posterior median for the proportion of level-$0$ agents in the \noun{All10} dataset according to the QLk model is 32\%, with a 95\% credible interval of [29\%,~35\%].
This is toward the high end of the range of previous estimates.
However, note that our estimate for QLk is very similar to the fitted estimate of \citet{burchardi14}, and comfortably within the range that they estimated by directly evaluating subjects' elicited strategies in a single game.
%
\Store{q:lk-identifiability}{
According to the Lk model, the posterior median for the proportion of level-$0$ agents in \noun{All10} is 18\%.  However, the Lk model suffers from an identifiability problem, in that there is no way to
distinguish uniform noise that is introduced by the uniform error structure from uniform noise introduced by level-$0$ agents.  This results in a very wide 95\% credible interval of [1\%,~42\%].
}

In contrast to our estimates, the number of level-$0$ agents in the population is typically assumed to be negligible in studies that use an iterative model of behavior.  Indeed, some studies \citeeg{crawford07} fix the number of level-$0$ agents to be 0.  Thus, one possible interpretation of our higher estimates of level-$0$ agents is as evidence of a misspecified model. For example, Poisson-CH uses level-$0$ agents as the only source of noisy responses.  However, we estimated substantial proportions of level-$0$ agents even for models (Lk and QLk) that include explicit error structures.  We thus believe that the alternative---that nonstrategic behavior occurs at a substantial frequency---must be taken seriously.

\section{Model Variations}
\label{sec:variations}
QLk makes various modeling assumptions that may seem arbitrary. For example, is it the right choice to model exactly two cognitive levels? And, is it really necessary to model the fact that agents at one level might be incorrect about the precision of the level below them? We now investigate these and other such questions, considering a family of models that systematically vary the assumptions underlying QLk. In the end, we identify a simpler model that dominated QLk on our data.

\begin{table}[tb]\centering
  \caption[Model variations with prediction performance]{
      Model variations with prediction performance on the \noun{All10} dataset.  The models with max level of $*$ used a Poisson distribution.  Models are named according to precision beliefs, precision homogeneity, population beliefs, and type of level distribution.  E.g., \texttt{ah-QCH3} is the model with accurate precision beliefs, homogeneous precisions, cognitive hierarchy population beliefs, and a discrete distribution over levels $0$--$3$.}
  ~\\

  {\scriptsize{}
  \begin{tabular}{rccllcr}
    \toprule
        Name           & \parbox[b]{0.2in}{Max Level}  & \parbox[b]{0.5in}{\centering Population Beliefs} &  \parbox[b]{0.5in}{Precision Beliefs} & Precisions & Parameters & \parbox[b]{0.62in}{\centering\mbox{Log likelihood} vs.\ u.a.r.}\\
    \midrule
            \texttt{QLk1} & 1   &    n/a & n/a        & n/a          &  2  &  $87.37   \pm  1.04$  \\
         \texttt{gi-QLk2} & 2   &    Lk  &  general   & inhomo.      &  5  &  $108.66  \pm  0.56$  \\
         \texttt{ai-QLk2} & 2   &    Lk  & accurate   & inhomo.      &  4  &  $103.33  \pm  1.75$  \\
         \texttt{gh-QLk2} & 2   &    Lk  & general    & homo.        &  4  &  $107.96  \pm  0.46$  \\
         \texttt{ah-QLk2} & 2   &    Lk  & accurate   & homo.        &  3  &  $104.84  \pm  0.58$  \\
         \texttt{gi-QCH2} & 2   &    CH  & general    & inhomo.      &  5  &  $107.78  \pm  0.88$  \\
         \texttt{ai-QCH2} & 2   &    CH  & accurate   & inhomo.      &  4  &  $106.76  \pm  0.92$  \\
         \texttt{gh-QCH2} & 2   &    CH  & general    & homo.        &  4  &  $109.43  \pm  0.58$  \\
         \texttt{ah-QCH2} & 2   &    CH  & accurate   & homo.        &  3  &  $106.67  \pm  0.41$  \\
         \texttt{gi-QLk3} & 3   &    Lk  & general    & inhomo.      &  9  &  $113.17  \pm  1.46$  \\
         \texttt{ai-QLk3} & 3   &    Lk  & accurate   & inhomo.      &  6  &  $109.62  \pm  1.21$  \\
         \texttt{gh-QLk3} & 3   &    Lk  & general    & homo.        &  7  &  $113.48  \pm  1.46$  \\
         \texttt{ah-QLk3} & 3   &    Lk  & accurate   & homo.        &  4  &  $107.12  \pm  0.46$  \\
         \texttt{gi-QCH3} &  3   &  CH  &  general  &  inhomo.          &  10  &  $113.01  \pm  0.93$  \\
         \texttt{ai-QCH3} & 3   &    CH  & accurate   & inhomo.      &  6  &  $111.34  \pm  0.59$  \\
         \texttt{gh-QCH3} & 3   &    CH  & general    & homo.        &  8  &  $113.08  \pm  0.83$  \\
         \texttt{ah-QCH3} &  3 &  CH  &  accurate &  homo.              &  4  &  $110.42  \pm  0.46$  \\
         \texttt{ai-QLk4} & 4   &    Lk  & accurate   & inhomo.      &  8  &  $110.30  \pm  0.93$  \\
         \texttt{ah-QLk4} & 4   &    Lk  & accurate   & homo.        &  5  &  $106.63  \pm  0.71$  \\
         \texttt{ah-QLk5} & 5   &    Lk  & accurate   & homo.        &  6  &  $107.18  \pm  0.57$  \\
         \texttt{ah-QLk6} & 6   &    Lk  & accurate   & homo.        &  7  &  $106.57  \pm  0.68$  \\
         \texttt{ah-QLk7} & 7   &    Lk  & accurate   & homo.        &  8  &  $106.50  \pm  0.69$  \\
         \texttt{ah-QLkp} & *   &    Lk  & accurate   & homo.        &  2  &  $106.89  \pm  0.28$  \\
         \texttt{ai-QCH4} & 4   &    CH  & accurate   & inhomo.      &  8  &  $111.54  \pm  0.62$  \\
         \texttt{ah-QCH4} & 4   &    CH  & accurate   & homo.        &  5  &  $110.88  \pm  0.33$  \\
         \texttt{ah-QCH5} & 5   &    CH  & accurate   & homo.        &  6  &  $111.22  \pm  0.39$  \\
         \texttt{ah-QCH6} & 6   &    CH  & accurate   & homo.        &  7  &  $111.26  \pm  0.44$  \\
         \texttt{ah-QCH7} & 7   &    CH  & accurate   & homo.        &  8  &  $111.42  \pm  0.41$  \\
         \texttt{ah-QCHp} & *   &    CH  & accurate   & homo.        &  2  &  $110.48  \pm  0.25$  \\
    \bottomrule
  \end{tabular}
  }
  \label{tbl:hypercube}
\end{table}

More specifically, we considered four different axes along with the QLk model could be modified. First, QLk assumes a maximum level of 2; we considered maximum levels of 1 and 3 as well.  Second, QLk assumes \emph{inhomogeneous precisions} in that it allows each level to have a different precision; we varied this by also considering \emph{homogeneous precision} models.  Third, QLk allows \emph{general precision beliefs} that can differ from lower-level agents' true precisions; we also constructed models that make the simplifying assumption that all agents have \emph{accurate precision beliefs} about lower-level agents.\footnote{This is in the same spirit as the simplifying assumption made in cognitive hierarchy models that agents have accurate beliefs about the proportions of lower-level agents.}
Finally, in addition to \emph{Lk} beliefs, where all other agents are assumed by a level-$k$ agent to be level-$(k-1)$, we also constructed models with \emph{CH} beliefs, where agents believe that the population consists of the true, truncated distribution over the lower levels.
We evaluated each combination of axis values; the 17 resulting models\footnote{When the maximum level is 1, all combinations of the other axes yield identical predictions.  Therefore there are only 17 models instead of $3(2^3)=24$.} are listed in the top part of Table~\ref{tbl:hypercube}.
In addition to the 17 exhaustive axis combinations for models with maximum levels in $\{1,2,3\}$, we also evaluated (1) 12 additional axis combinations that have higher maximum levels and 8 parameters or fewer: \texttt{ai-QCH4} and \texttt{ai-QLk4}; \texttt{ah-QCH} and \texttt{ah-QLk} variations with maximum levels in $\{4,5,6,7\}$; and (2) \texttt{ah-QCH} and \texttt{ah-QLk} variations that assume a Poisson distribution over the levels rather than using an explicit tabular distribution.\footnote{The \texttt{ah-QCHp} model is identical to the CH-QRE model of \citet{camerer16}.}  These additional models are listed in the bottom part of Table~\ref{tbl:hypercube}.

\subsection{Simplicity Versus Predictive Performance}
\label{sec:simplicity}

\begin{figure}[t]\centering
  \includegraphics[width=.9\textwidth]{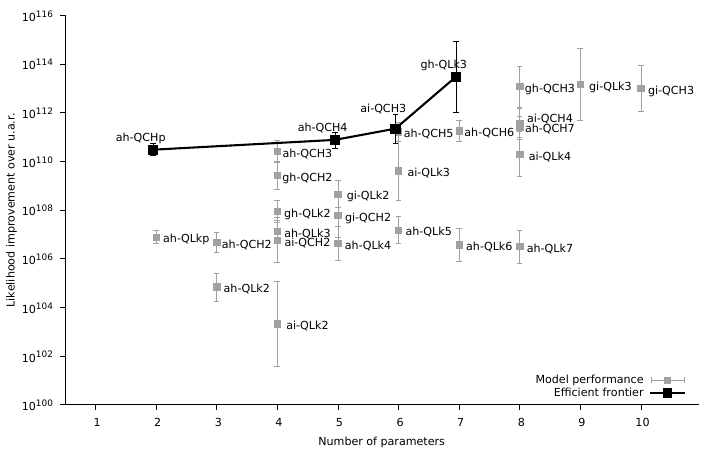}
  \vspace{-.5em}%
  \caption[Model simplicity vs.\ prediction performance]{
      Model simplicity vs.\ prediction performance on the \noun{All10} dataset.  \texttt{QLk1} is omitted because its far worse performance ($\sim 10^{87}$) distorts the figure's scale.}
  \label{fig:hypercube}
\end{figure}

We evaluated the predictive performance of each model on the \noun{All10} dataset using 10-fold cross-validation repeated 10 times, as in Section~\ref{sec:comparisons}.  The results are given in the last column of Table~\ref{tbl:hypercube} and plotted in Figure~\ref{fig:hypercube}.

All else being equal, a model with higher performance is more desirable, as is a model with fewer parameters.
We can plot an \emph{efficient frontier} of those models that achieved the best performance for a given number of parameters or fewer; see Figure~\ref{fig:hypercube}.
%
The original QLk model (\texttt{gi-QLk2}) is \emph{not} efficient in this sense; it is dominated by, e.g., \texttt{ah-QCH3}, which has both significantly better predictive performance and fewer parameters (because it restricts agents to homogeneous precisions and accurate beliefs).

There is a striking pattern among the efficient models with $6$ parameters or fewer: every such model has accurate precision beliefs, cognitive hierarchy population beliefs, and, with the exception of \texttt{ai-QCH3}, homogeneous precisions.  Furthermore, \texttt{ai-QCH3}'s performance was not significantly better than that of \texttt{ah-QCH5}, which did have homogeneous precisions.  This suggests that the most parsimonious way to model human behavior in normal-form games is to use a model of this form.

Adding flexibility by modeling general beliefs about precisions did improve performance; the four best-performing models all incorporated general precision beliefs.  However, these models also had much larger variance in their prediction performance on the test set.  This may indicate that the models are overly flexible, and hence prone to overfitting.

\subsection{Parameter Analysis of \texttt{ah-QCH} Models}
\label{sec:ah-qch-results}

\begin{figure}[tb]\centering
  \includegraphics[width=.95\textwidth]{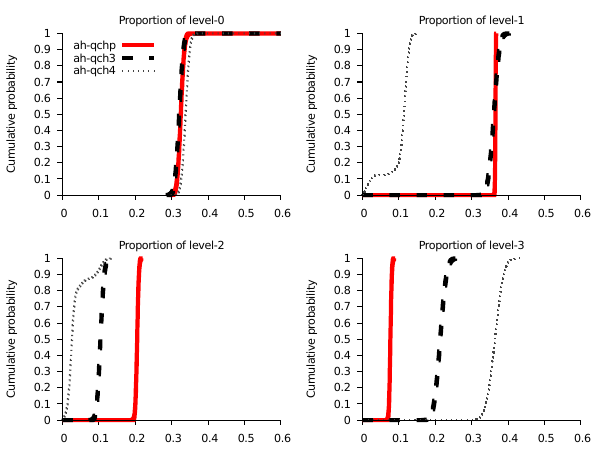}
  \caption[Posterior distributions for \texttt{ah-QCH3} and \texttt{ah-QCH4}]{
      Marginal cumulative posterior distributions for the level proportion parameters ($\alpha_0,\alpha_1,\alpha_2,\alpha_3$) of the \texttt{ah-QCHp}, \texttt{ah-QCH3}, and \texttt{ah-QCH4} models on \noun{All10}.  Solid lines are \texttt{ah-QCHp}; dashed lines are \texttt{ah-QCH3}; dotted lines are \texttt{ah-QCH4}.
      All $\alpha$ values are defined implicitly by the $\tau$ parameter for \texttt{ah-QCHp}.  For the other models,
      $\alpha_0$ is defined implicitly by $\alpha_1, \alpha_2, \alpha_3,$, and (for \texttt{ah-QCH4}) $\alpha_4$.}
  \label{fig:cdf-ah-qch}
\end{figure}

In this section we examine the marginal posterior distributions of two models from the accurate, homogeneous QCH family (see Figure~\ref{fig:cdf-ah-qch}).  We computed the posterior distribution of the models' parameters using the procedure described in Sections~\ref{sec:posterior-estimation} and~\ref{sec:parameters}.
%
The posterior distribution for the precision parameter $\lambda$ was concentrated around $0.20$, somewhat greater than the QLk model's estimate for $\lambda_1$.  This suggests that QLk's much lower estimate for $\lambda_{1(2)}$ may indeed have been the closest that the model could get to having the level-$2$ agents best respond to a mixture of level-$0$ and level-$1$ agents (as in cognitive hierarchy).

Our robust finding in
Sections~\ref{sec:level0} and~\ref{sec:qlk-results}
of a large proportion of level-$0$ agents was confirmed by these models as well.  Indeed, the number of level-$0$ agents was nearly the only point of close agreement between all three models with respect to the distribution of levels.


\section{Related Work}
\label{sec:perf-related}
Our work has been motivated by the question, ``What model is best for predicting human behavior in general, simultaneous-move games?''  Before beginning our study, we conducted an exhaustive literature survey to determine the extent to which this question had already been answered. Specifically, we used Google Scholar to identify all (1805) citations to the papers introducing the QRE, CH, Lk, NI, and QLk models \citep{mckelvey95,camerer04cognitive,costagomes01,nagel95,goeree04,stahl94}, and manually checked every reference.
We discarded superficial citations, papers that simply applied one of the models to an application domain, and papers that studied repeated games. This left us with a total of 24 papers, including the six with which we began, which we summarize in Table~\ref{tbl:big-table}.
%
Overall, we found no paper that compared the predictive performance of all six models. Indeed, there are two senses in which the literature focuses on different issues. First, it appears to be more concerned with \emph{explaining} behavior than with \emph{predicting} it.  Thus, comparisons of out-of-sample prediction performance were rare.  Here we describe the only exceptions that we found:
\begin{itemize}
	\item \citet{stahl95} evaluated prediction performance on 3 games using parameters fit from the other games;
	\item \citet{morgan02} and \citet{hahn10} evaluated prediction performance using held-out test data;
	\item \citet{camerer04cognitive} and \citet{chong05} computed likelihoods on each individual game in their datasets after using models fit to the $n-1$ remaining games;
	\item \citet{crawford07fatal} compared the performance of two models by training each model on each game in their dataset individually, and then evaluating the performance of each of these $n$ trained models on each of the $n-1$ other individual games; and
	\item \citet{camerer16} evaluated the performance of QRE and cognitive hierarchy variants on one experimental treatment using parameters estimated on two separate experimental treatments. 
\end{itemize}
Second, most of the papers compared a single one of the five models (often with variations) to Nash equilibrium. Indeed, only nine of the 24 studies (see the bottom portion of Table~\ref{tbl:big-table}) compared more than one of the six key models, and none of these considered QLk. Only three of these studies explicitly compared the prediction performance of more than one of the six models \citep{chong05,crawford07fatal,camerer16}; the remaining six performed comparisons in terms of training set fit \citep{camerer01,goeree04,costagomes08,costagomes09,rogers09,breitmoser12}.

\citet{rogers09} proposed a unifying framework that generalizes both Poisson-CH and QRE, and compared the fit of several variations within this framework. Notably, their framework allows for quantal response within a cognitive hierarchy model.  Their work is thus similar to our own search over a system of QLk variants in Section~\ref{sec:variations}, but there are several differences.  First, we compared out-of-sample prediction performance, not in-sample fit.  Second, \citeauthor{rogers09} restricted the distributions of types to be grid, uniform, or Poisson distributions, whereas we considered unconstrained discrete distributions over levels.  Third, they required different types to have different precisions, while we did not.  Finally, we considered level-$k$ beliefs as well as cognitive hierarchy beliefs, whereas they considered only cognitive hierarchy belief models.

One line of work in computer science also meets our criteria of predicting action choices and modeling human behavior \citep{altman06}. This approach learns association rules between agents' actions in different games to predict how an agent will play based on its actions in earlier games.  We did not consider this approach in our study, as it requires data that identifies agents across games, and cannot make predictions for games that are not in the training dataset.

\begin{table}[t]\centering
\normalfont%
\caption[Existing work in model comparison]{Existing work in model comparison. `f' indicates comparison of training sample fit only; `t' indicates statistical tests of training sample performance; `p' indicates evaluation of out-of-sample prediction performance.}%
\begin{tabular}{lcccccc}
\toprule
Paper &
  Nash &
  QLk&
  Lk&
  CH&
  NI&
  QRE\\
\midrule
  \citet{stahl94} & t & t &  &  &  & \\
  \citet{mckelvey95} & f &  &  &  & & f \\
  \citet{stahl95} & f & p &  &  & &  \\
  \citet{costagomes98} & f &  & f &  &  & \\
  \citet{haruvy99} &  & t &  &  &  & \\
  \citet{costagomes01} & f &  & f &  &  & \\
  \citet{haruvy01} &   & t &  &  &  & \\
  \citet{morgan02} & f &  &  &  & & p \\
  \citet{weizsacker03} & t &  &  &  & & t \\
  \citet{camerer04cognitive} & f &  &  & p &  & \\
  \citet{costagomes06} & f &  & f &  & & \\
  \citet{stahl08} &  & t &  &  &  &  \\
  \citet{reybiel09} & t &  & t &  &  & \\
  \citet{georganas15} & f &  & f &  &  &\\
  \citet{hahn10} &  &  &  & p &  & \\
\midrule
  \citet{camerer01} &  &  &  & f & & f \\
  \citet{goeree04}  & f &  &  &   & f & f  \\
  \citet{chong05} & f &  &  & p & & p \\
  \citet{crawford07fatal} & p &  & p & & & p \\
  \citet{costagomes08} & f &  & f &   & f & f \\
  \citet{costagomes09} & f &  & f & f & f & f \\
  \citet{rogers09} & f &  &  & f & & f \\
  \citet{camerer16} &  &  &  & p & & p \\
  \citet{breitmoser12} &  &  & t & t & t & t \\
\bottomrule
\end{tabular}
\label{tbl:big-table}%
\end{table}

\section{Conclusions}
\label{sec:conclusions}

To our knowledge, ours is the first study to address the question of which existing behavioral model---QRE, level-$k$, cognitive hierarchy, noisy introspection, or quantal level-$k$ behavioral models---is best suited to predicting unseen human initial play of normal-form games.  We explored the prediction performance of these models, along with several modifications.
%
We found that bounded iterated reasoning and cost-proportional errors are both valuable ingredients in a predictive model of human game theoretic behavior:  the best-performing model that we studied (QLk) combines both of these elements.
%
We believe that iterative reasoning describes an actual cognitive process. The situation is less clear with cost-proportional errors: they may likewise describe human reasoning, or they may simply be a closer approximation to human behavior than the usual uniform error specification.

Bayesian parameter analysis is a valuable technique for investigating the behavior and properties of models, particularly because it is able to make quantitative recommendations for parameter values. We showed how Bayesian parameter analysis can be applied to derive concrete recommendations for the use of Poisson-CH, differing substantially from widely cited advice in the literature.

QLk (\texttt{gi-qlk2}) provides substantial flexibility in specifying the beliefs and precisions of different types of agents.  We found that this flexibility tends to hurt generalization performance more than it helps.
In a systematic search of model variations, we identified a new model family (the accurate precision belief, homogeneous-precision QCH models) that contained the efficient (or nearly-efficient) model for every number of parameters smaller than $7$.  Based on further analysis of this model family, we identified a model, Poisson-QCH, that offers excellent generalization performance with only two parameters.

\subsection{Recommendations}
\label{sec:recommendations}
\paragraph{Methodology}
In this work we have focused exclusively on prediction performance.
One might wonder whether there is any practical difference between in-sample fit and out-of-sample prediction performance.
It turns out that the ranking of a model's performance within a dataset was identical in the test and training sets only 45\% of the time, despite the low dimensionality of the models that we considered.  The average difference between a model's rank by test performance and its rank by training performance was 1.5.
The \texttt{ai-qlk4} model was an especially notable example, having the $5^\textup{th}$-highest training performance but only the $14^\textup{th}$-highest test performance.

We thus conclude that there is no substitute for evaluating a model on held-out test data.
We recommend the use of 10-fold cross-validation, repeated 10 times with a different random partition over games on each repetition, as described in Section~\ref{sec:methods}.
However, we recognize that this process is computationally intensive, as it requires each model to be fit 100 times.
If computation time is a major constraint, we recommend a single round of 10-fold cross-validation, or even a single round of 4-fold cross-validation; this still gives an unbiased estimate of prediction performance, albeit without error bars.

The log-likelihood performance measure has some problematic features: it is not comparable between datasets, and its units do not have an especially natural interpretation.
Nevertheless, it is the most appropriate performance measure for predictive behavioral models of which we are aware, especially when normalized against a baseline such as the performance of uniform predictions.

\paragraph{Models}
\label{sec:model-recommendation}
\Store{q:frontier}{%
Section~\ref{sec:variations} analyzes an ``efficient frontier'' of models, each of which represent a different tradeoff between performance and parsimony (and hence robustness).  The Poisson-QCH model (\texttt{ah-QCHp}) is attractive for being low-variance and reasonably performant, whereas \texttt{gh-QLk3} has the highest expected performance but also the highest variance of any model, and a more difficult-to-interpret parameter structure.
}

We recommend the use of the Poisson-QCH model for the prediction of human strategic behavior in unrepeated, simultaneous-move games.\footnote{Equilibrium-based theories may have more of a role to play in the repeated setting, where agents have a chance to converge to equilibrium (although see \citet{frey13} for evidence against convergence in a repeated setting).}  
The median posterior parameters for the \noun{All10} dataset were $\lambda=0.20, \tau=1.12$.\footnote{This suggested value for $\tau$ may seem superficially similar to the value $\tau=1.5$ suggested by \citet{camerer04cognitive} for Poisson-CH.  However, they differ quite meaningfully, as $\tau=1.12$ implies that $33$\% of the population are level-$0$, whereas $\tau=1.5$ implies that only $22$\% are level-$0$.}
These settings may be a good starting point for applications, although we note that application-specific fits are always preferable due to behavioral variation across subject populations.

\subsection{Further Directions}
Our parameter estimates for all of the iterative models included a substantial proportion of level-$0$ agents.  The level-$0$ model is important for predicting the behavior of all agents in an iterative model; both the level-$0$ agents themselves, and the higher-level agents whose behavior is grounded in a model of level-$0$ behavior.  In ongoing work, we are investigating richer specifications of level-$0$ behavior, which allow for significant performance improvements \citep{wright14level}.

\Store{q:endo-preview}{
Our approach of fitting the parameters of an iterative model in one set of games and then using these parameters to make predictions in distinct games implicitly assumes that the distribution of beliefs in the population is constant across different games.  In several studies, experimental subjects do exhibit surprising stability \citep{stahl94,stahl95,costagomes01,polonio15} or convergence \citep{breitmoser14} in their apparent levels of reasoning.
However, it also seems reasonable to suppose that players' depths of reasoning would be influenced by the structure of the game.  In ongoing work, we are investigating ways to model such endogeneous reasoning steps.
}

\section*{Acknowledgements}
This work was funded in part by the Natural Sciences and Engineering Research Council of Canada.
It was completed in part while the authors were visiting the Simons Institute for the Theory of Computing.
We thank several anonymous reviewers and editors for many helpful comments that have significantly improved the paper.

\bibliography{journal-prediction}
\bibliographystyle{apalike}

\appendix
\section{Likelihood Derivation}
\label{apx:likelihood}

The likelihood of a single datapoint $d_{ij} = (G_i, a_{ij})$ is
\[ \Pr(d_{ij} \| \theta) = \Pr(G_i, a_{ij} \| \theta). \]
By the chain rule of probabilities, this\footnote{To those unfamiliar with Bayesian analysis, quantities such as $\Pr(\D)$, $\Pr(G_i)$, and $\Pr(G_i \| \theta)$ may seem difficult to interpret or even nonsensical.  It is common practice in Bayesian statistics to assign probabilities to any quantity that can vary, such as the games under consideration or the complete dataset that has been observed.  Regardless of how they are interpreted, these quantities all turn out to be constant with respect to $\theta$, and so have no influence on the outcome of the analysis.}
is equivalent to
\[ \Pr(d_{ij} \| \theta) = \Pr(a_{ij} \| G_i, \theta)\Pr(G_i \| \theta),\]
and by independence of $G$ and $\theta$ we have
\begin{equation}
  \Pr(d_{ij} \| \theta) = \Pr(a_{ij} \| G_i, \theta)\Pr(G_i).
\end{equation}
The datapoints are independent, so the likelihood of the dataset is just the product of the likelihoods of the datapoints,
\begin{equation}
\Pr(\D \| \theta) = \prod_{i=1}^I\prod_{j=1}^{J_i} \Pr(a_{ij} \| G_i, \theta)\Pr(G_i).\label{eq:likelihood}
\end{equation}
The probabilities $\Pr(G_i)$ are constant with respect to $\theta$, and can therefore be disregarded when maximizing the likelihood:
\[  \argmax_{\theta} \Pr(\D \| \theta) = \argmax_{\theta} \prod_{i=1}^I\prod_{j=1}^{J_i} \Pr(a_{ij} \| G_i, \theta). \]

\section{Posterior Distribution Derivation}
\label{apx:posterior}

We derive an expression for the posterior distribution
$\Pr(\theta \| \D)$ by applying Bayes' rule, where $p_0(\theta)$ is the prior distribution:
\begin{equation}
  \Pr(\theta \| \D) = \frac{p_0(\theta)\Pr(\D \| \theta)}{\Pr(\D)}. \label{eq:posterior}
\end{equation}
Substituting in Equation \eqref{eq:likelihood}, which gave an expression for the likelihood of the dataset $\Pr(\D \| \theta)$, we obtain
\begin{equation}
  \Pr(\theta \| \D) = \frac{p_0(\theta)\prod_{i=1}^I\prod_{j=1}^{J_i} \Pr(a_{ij} \| G_i, \theta)\Pr(G_i)}{\Pr(\D)}.
\end{equation}
In practice $\Pr(G_i)$ and $\Pr(\D)$ are constants, and so can be ignored:
\begin{equation}
  \Pr(\theta \| \D) \propto p_0(\theta)\prod_{i=1}^I\prod_{j=1}^{J_i} \Pr(a_{ij} \| G_i, \theta).
\end{equation}
Note that by commutativity of multiplication, this is equivalent to performing iterative Bayesian updates one datapoint at a time.  Therefore, iteratively updating this posterior neither over- nor underprivileges later datapoints.


\section{Dataset Composition}
\label{apx:data-composition}
As we saw in the case of \noun{GH01}, model performance was sensitive to choices made by the authors of our various datasets about which games to study. One way to control for such choices is to partition our set of games according to important game properties, and to evaluate model performance in each partition.  In this appendix we describe such an analysis.

Overall, our datasets spanned 142 games. The vast majority of these games are matrix games, deliberately lacking inherent meaning in order to avoid framing effects.\footnote{Indeed, some studies \citeeg{rogers09} even avoided focal payoffs like 0~and~100.}  For the most part, these games were chosen to vary according to dominance solvability and equilibrium structure.
%
In particular, most dataset authors were concerned with (1) whether a game could be solved by iterated removal of dominated strategies (either strict or weak) and with how many steps of iteration were required; and (2) the number and type of Nash equilibria that each game possesses.\footnote{There were two exceptions. The first was \citet{goeree01}, who chose games that had both equilibria that human subjects find intuitive and strategically equivalent variations of these games whose equilibria human subjects find counterintuitive. The second exception was \citet{cooper03}, whose normal form games were based on an exhaustive enumeration of the payoff orderings possible in generic $2$-player, $2$-action extensive-form games.}

\begin{table}[t]\centering
  \caption[Datasets conditioned on various game features]{
    Datasets conditioned on various game features.  The column headed ``games'' indicates how many games of the full dataset met the criterion, and the column headed ``$n$'' indicates how many observations each feature-based dataset contained. Observe that the game features are not all mutually exclusive, and so the ``games'' column does not sum to 142.}
  \begin{tabular}{lp{2.5in}rr}
    \toprule
    Name & Description & Games & $n$ \\
    \midrule
    \noun{D1}         & Weak dominance solvable in one round    &  2 &  748  \\
    \noun{D1s}        & Strict dominance solvable in one round  &  0 &    0  \\
    \noun{D2}         & Weak dominance solvable in two rounds   & 38 & 5058 \\
    \noun{D2s}        & Strict dominance solvable in two rounds & 23 & 2000 \\
    \noun{DS}         & Weak dominance solvable    & 52 & 6470 \\
    \noun{DSs}        & Strict dominance solvable  & 35 & 3312 \\
    \noun{ND}         & Not dominance solvable                  & 90 & 7393 \\
    \midrule
    \noun{PSNE1}      & Single Nash equilibrium, which is pure  & 51 & 4687 \\
    \noun{MSNE1}      & Single Nash equilibrium, which is mixed & 21 & 1387 \\
    \noun{Multi-Eqm}  & Multiple Nash equilibria                & 70 & 7789 \\
    \bottomrule
  \end{tabular}
  \label{tbl:sifted-datasets}
\end{table}

\begin{figure}[t]\centering
  \includegraphics{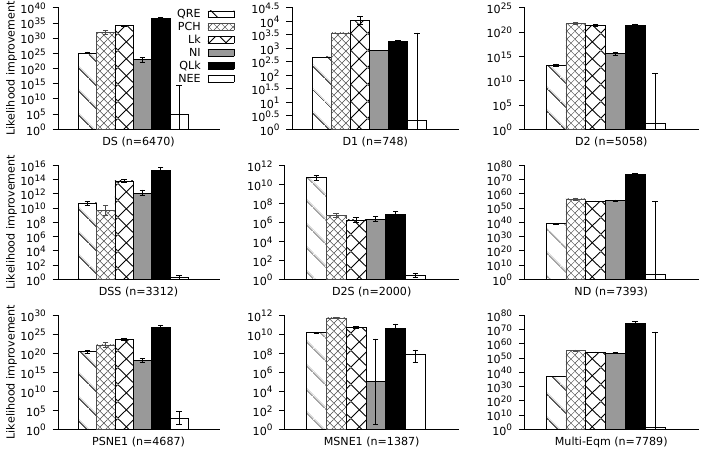}
  \caption[Likelihoods on feature-based datasets]{
    Average likelihood ratios of model predictions to random predictions, with 95\% confidence intervals, on feature-based datasets.
    For NEE the main bar shows performance averaged over all equilibria and error bars show post-hoc upper and lower bounds on equilibrium performance.}
  \label{fig:sifted-mle}
\end{figure}

We thus constructed subsets of the full dataset based on their dominance solvability and the nature of their Nash equilibria, as described in Table~\ref{tbl:sifted-datasets}.\footnote{\label{fn:balance}
As Table~\ref{tbl:sifted-datasets} shows, there was some variance in the number of games and observations among the different partitions.
The results presented in this appendix indicate that this variance was likely not a major determinant of our overall results.}
%
We computed cross-validated MLE fits for each model on each of the feature-based datasets of Table~\ref{tbl:sifted-datasets}.  The results are summarized in Figure~\ref{fig:sifted-mle}.  In two respects, the results across the feature-based datasets mirror the results of Section~\ref{sec:bgt-comparison} and Section~\ref{sec:nash-comparison}.  First, QLk significantly outperformed the other behavioral models on the majority of datasets; the exceptions were \noun{D1}, \noun{D2}, and \noun{D2s} (but not \noun{DS}); and \noun{MSNE1}.  
%
Second, a majority of behavioral models significantly outperformed NEE in all but three datasets: \noun{D1}, \noun{ND} and \noun{Multi-eqm}.
In these three datasets, the upper and lower bounds on NEE's performance contained the performance of either two or all three of the single-factor behavioral models (but not necessarily QLk).
It is unsurprising that NEE's upper and lower bounds were widely separated on the \noun{Multi-eqm} dataset, since the more equilibria a game has, the more variation there can be in these equilibria's post-hoc performance; NEE's strong best-case performance on this dataset should similarly reflect this variation.
It turns out that 55 of the 90 games (and 4731 of the 7393 observations) in the \noun{ND} dataset are from the \noun{Multi-eqm} dataset, which likely explains NEE's high upper bound in that dataset as well.
Indeed, this analysis helps to explain some of our previous observations about the \noun{GH01} dataset.  NEE contains all other models in its performance bounds in this dataset, and in addition to the fact that half the dataset's games (the ``treasure'' treatments) that were chosen for consistency with Nash equilibrium, some of the other games (the ``contradiction'' treatments) turn out to have multiple equilibria. Overall, the overlap between \noun{GH01} and \noun{Multi-eqm} is 5 games out of 10 and 250 observations out of 500.

Unlike in the per-dataset comparisons of Section~\ref{sec:bgt-comparison}, both of our iterative single-factor models (Poisson-CH and Lk) significantly outperformed QRE in almost every feature-based dataset, with D2S and DSS as the only exceptions; in D2S, QRE outperformed all other models, and in DSS QRE was significantly outperformed by Lk but not by Poisson-CH.
One possible explanation is that the filtering features are all biased toward iterative models.  However, it seems unlikely that, e.g., \emph{both} dominance-solvability and dominance-nonsolvability are biased toward iterative models.  Another possibility is that iterative models are a better model of human behavior, but the cost-proportional error model of QRE is sufficiently superior to the respectively simple and non-existent error models of Lk and Poisson-CH that it outperforms on many datasets that mix game types.  However, we observed no straightforward relationship between the different proportions of dominance-solvable and non-dominance-solvable games in a source dataset and the relative performance of Lk/Poisson-CH and QRE.

\end{document}